\documentclass[fp,twocolumn]{jpsj3}

\usepackage{txfonts}

\usepackage{amsmath}
\usepackage{amsfonts}
\usepackage{bm}
\usepackage{graphicx}

\title{Can we swim in superfluids?: Numerical demonstration of
self-propulsion in a Bose-Einstein condensate}

\author{Hiroki Saito}
\inst{Department of Engineering Science, University of
Electro-Communications, Tokyo 182-8585, Japan}

\abst{It is numerically investigated whether a deformable object can
propel itself in a superfluid.
Articulated bodies and multi-component condensates are examined as
swimmers.
An articulated two-body swimmer cannot obtain locomotion without emitting
excitations.
More flexible swimmers can do so without the need to excite waves.
}

\begin{document}
\maketitle

\section{Introduction}

Superfluids have no viscosity.
When an object is moved in a superfluid at a subcritical constant
velocity, it experiences no drag force.
It would seem, therefore, that the arms and legs of a swimmer will be
unable to generate propulsion in a superfluid.
Can we swim in such a fluid?
In other words, is the self-propulsion of a deformable object in a
superfluid possible?

The present paper shows that the answer to this question is yes.
Let us first consider a case in which the velocity of the flow around the
swimmer is much slower than the velocity of sound.
In this case, a superfluid may be regarded as an incompressible ideal
fluid.
Several authors have shown that the self-propulsion of a deformable body
is possible even in an incompressible ideal
fluid~\cite{Benjamin,Saffman,Miloh,Kanso,Chambrion}, and therefore, we
expect that self-propulsion is also possible in a superfluid.
When the swimmer deforms quickly, on the other hand, the superfluid cannot
be regarded as incompressible, and waves and quantized vortices are
generated.
Since these excitations are able to carry momentum, the swimmer can obtain
thrust by releasing them.
The aim of the present paper is to verify these speculations.

The dynamics of objects immersed in superfluids have been studied in
various situations, such as using particles to visualize superfluid
flow~\cite{Donnelly,Zhang}, using oscillating spheres~\cite{Jager} and
wires~\cite{Goto} to disturb superfluid helium, and considering the
behavior of an ion~\cite{Zipkes} in a gaseous Bose-Einstein condensate
(BEC).
However, the self-propulsion of an object in a superfluid has not been
studied.

In the present paper, we use the mean-field Gross-Pitaevskii (GP) theory
to numerically demonstrate that deformable objects can swim in a
two-dimensional (2D) superfluid.
Two kinds of swimmers are examined:
articulated bodies with movable joints and multicomponent BECs.
For the latter type, the swimmer is also a BEC and can be deformed by
changing the interactions between atoms.
Since a superfluid has no viscosity, one may think that the swimmer would
behave as if it had a high Reynolds number.
However, the condition for self-propulsion in a superfluid is shown to
be similar to that for a swimmer with a low Reynolds number in a viscous
fluid; this is reminiscent of Purcell's ``scallop
theorem''~\cite{Purcell}.

This paper is organized as follows.
Section~\ref{s:link} investigates the dynamics of articulated bodies in a
BEC.
Section~\ref{s:multi} studies the dynamics of multicomponent BECs.
Section~\ref{s:conc} presents the conclusions of our study.

\section{Articulated bodies in a superfluid}
\label{s:link}

We consider the dynamics of a superfluid of atoms with mass $m$ in the
mean-field approximation.
The dynamics of the macroscopic wave function $\psi$ are described by the
GP equation
\begin{equation} \label{GP}
i \hbar \frac{\partial\psi}{\partial t} = -\frac{\hbar^2}{2m} \nabla^2
\psi + V \psi + g |\psi|^2 \psi,
\end{equation}
where $V$ is the potential produced by a swimmer.
For simplicity, we will consider a 2D space, which can be
realized by tightly confining the system in the $z$ direction.
Assuming the form of the wave function is $\psi_{\rm 3D}(x, y, z) =
\psi(x, y) \phi(z)$, the effective interaction coefficient $g$ in
Eq.~(\ref{GP}) is given by $g = 4 \pi \hbar^2 a \int |\phi|^4 dz / m$,
where $a$ is the $s$-wave scattering length.

In this section, we consider swimmers that consist of elliptic bodies, as
illustrated in Fig.~\ref{f:swimmers}.
The elliptic bodies are connected at joints located on the long axes of
the ellipses.
The distance between the centers of the ellipses and the joints is $d$.
The angles $\bm{\chi}$ of the joints are controlled variables.
To reduce numerical errors arising from the rigid boundary condition at
the surface of the bodies, a ``soft'' boundary condition is adopted.
For the $j$th elliptic body centered at $(x_j, y_j)$ and for which the
angle between its long axis and the $x$ axis is $\theta_j$, the potential
is given by
\begin{equation} \label{ellpot}
v_j(x, y) = \mu_0 \exp\{ f_{\rm pot} [1 - (\xi_j / \alpha)^2
- (\eta_j / \beta)^2] \},
\end{equation} 
where $\xi_j = (x - x_j) \cos\theta_j + (y - y_j) \sin\theta_j$,
$\eta_j = -(x - x_j) \sin\theta_j + (y - y_j) \cos\theta_j$, and $\mu_0$
is the chemical potential far from the swimmer.
The potential in Eq.~(\ref{ellpot}) is $v_j = \mu_0$ at an ellipse $(\xi_j
/ \alpha)^2 + (\eta_j / \beta)^2 = 1$, which corresponds to the surface of
the body.
The parameter $f_{\rm pot}$ in Eq.~(\ref{ellpot}) determines the sharpness
of the surface; the Gaussian potential rises steeply for a large
$f_{\rm pot}$.
The results are insensitive to the value of $f_{\rm pot}$, and we set
$f_{\rm pot} = 1$ in the following calculations.
The potential $V$ in the GP equation is thus given by
\begin{equation} \label{V}
V(x, y) = \sum_j v_j(x, y).
\end{equation}

The Lagrangian for the swimmer is given by
\begin{equation} \label{L}
L = \sum_j \left[ \frac{M}{2} (\dot{x}_j^2 + \dot{y}_j^2)
+ \frac{I}{2} \dot{\theta}_j^2 \right]
- U_{\rm pot}(X, Y, \theta, \bm{\chi}; \psi),
\end{equation}
where $M$ and $I$ are the mass and the moment of inertia of the elliptic
body, respectively.
The interaction potential $U_{\rm pot}$ between the swimmer and superfluid
has the form
\begin{equation} \label{U}
U_{\rm pot}(X, Y, \theta, \bm{\chi}; \psi) = \int V |\psi|^2 d\bm{r},
\end{equation}
where $(X, Y)$ are coordinates of the center of mass (COM) of the
swimmer.
The potential $V$ depends on $(X, Y)$, $\theta$, and $\bm{\chi}$, through
Eq.~(\ref{ellpot}).
In the following, we assume $M = \pi \alpha \beta m n_0$ and $I = M
(\alpha^2 + \beta^2) / 4$, where $n_0$ is the density far from the
swimmer.

Equation~(\ref{GP}) and the equation of motion for the swimmer are solved
numerically, using the pseudospectral and fourth-order Runge-Kutta
methods.
The initial state of $\psi$ is prepared by the imaginary-time propagation
method, in which the swimmer is at rest in the initial form.
The density $|\psi|^2$ far from the swimmer is constant $n_0$, which gives
the characteristic length $\xi = \hbar / (m g n_0)^{1/2}$ and time $\tau =
\xi / v_s$, where $v_s = (g n_0 / m)^{1/2}$ is the velocity of sound.
The following quantities are fixed: $\alpha = 5 \xi$, $\beta = \xi$, and
$d = 7 \xi$.
A periodic boundary condition is imposed by the pseudospectral method.
The numerical space is taken to be sufficiently wide, and the boundary
does not affect the results if the motion of the swimmer is moderate.
In this case, the total momentum of the swimmer and surrounding fluid is
conserved.
When the swimmer moves violently, waves and vortices are released, and
these reach the boundary.
In that case, an artificial damping term $-\Gamma(\bm{r}) \hbar \partial
\psi / \partial t$ is added to the left-hand side of Eq.~(\ref{GP}),
where $\Gamma(\bm{r})$ is nonzero only near the boundary~\cite{Reeves}.
The artificial damping term can suppress the disturbances at the boundary
without affecting the dynamics of the swimmer.

\subsection{Two-body swimmer}
\label{s:2link}

\begin{figure}
\begin{center}
\includegraphics[width=8.5cm]{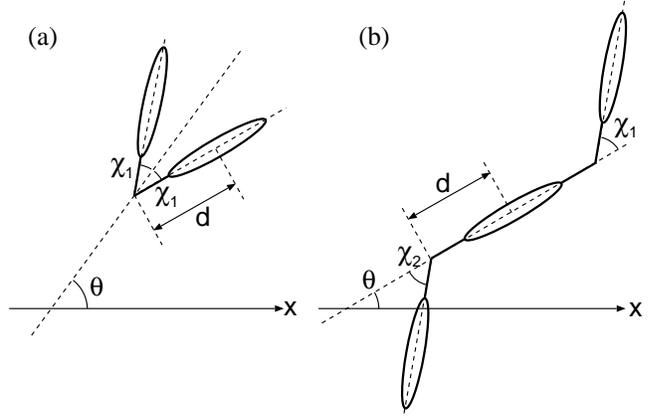}
\end{center}
\caption{
Schematic illustration of the two-body and three-body swimmers
used in Secs.~\ref{s:2link} and \ref{s:3link}, respectively.
The distance $d$ is constant between the center of the elliptic bodies and
the joints.
The angles $\chi_1$ and $\chi_2$ are controlled parameters.
}
\label{f:swimmers}
\end{figure}
We begin by considering a swimmer created from two elliptic bodies
connected by a joint, as shown in Fig.~\ref{f:swimmers}(a).
The state of the swimmer is specified by the COM coordinate $(X, Y)$,
direction $\theta$ of the symmetry axis, and their time derivatives $(\dot
X, \dot Y)$, $\dot\theta$.
The angle $\chi_1$ is the controlled variable, and the shape of the
swimmer has only one degree of freedom.

For the two-body swimmer, the Lagrangian in Eq.~(\ref{L}) is rewritten as
\begin{eqnarray} \label{L2body}
L & = & M (\dot{X}^2 + \dot{Y}^2) + \frac{M d^2}{2} \left[ \dot{\theta}^2
+ \dot{\chi}_1^2 - (\dot{\theta}^2 - \dot{\chi}_1^2) \cos 2\chi_1
\right]
\nonumber \\
& & + I (\dot{\theta}^2 + \dot{\chi}_1^2)
- U_{\rm pot}(X, Y, \theta, \chi_1; \psi).
\end{eqnarray}
The dynamics of the swimmer are thus described by the Euler-Lagrange
equations,
\begin{subequations} \label{EL2body}
\begin{eqnarray}
2 M \ddot{X} & = & -\frac{\partial U_{\rm pot}}{\partial X}, \\
2 M \ddot{Y} & = & -\frac{\partial U_{\rm pot}}{\partial Y}, \\
\left[ M d^2 (1 - \cos 2\chi_1) + 2I \right] \ddot{\theta} & = &
-\frac{\partial U_{\rm pot}}{\partial \theta} - 2 M d^2 \dot\theta
\dot{\chi}_1 \sin 2\chi_1. \nonumber \\
\end{eqnarray}
\end{subequations}
Equations~(\ref{GP}) and (\ref{EL2body}) determine the dynamics of the
system.

\begin{figure}
\begin{center}
\includegraphics[width=8.5cm]{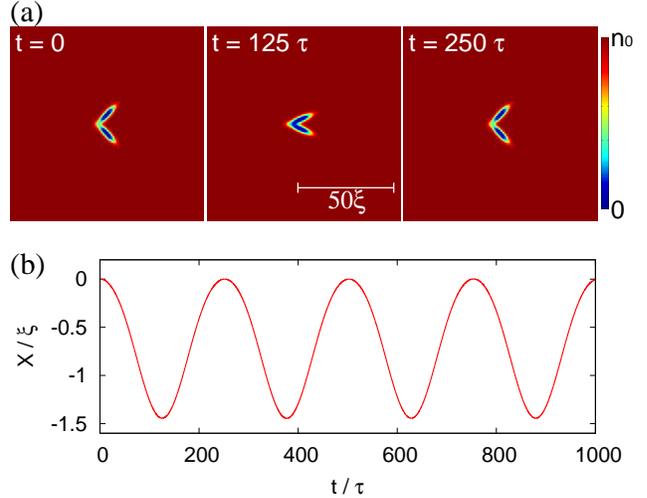}
\end{center}
\caption{
(Color online) Time evolution of (a) the density profile $|\psi|^2$, and
(b) the position $X$ of the COM.
The angle $\chi_1(t)$ is driven as in Eq.~(\ref{chi1}) with $A = \pi / 4$
and $\omega = 0.025 \tau^{-1}$.
The length and time are normalized by the healing length $\xi$ and $\tau =
\xi / v_s$, where $v_s$ is the sound velocity.
See the Supplemental Material for a movie of the dynamics shown in
(a)~\cite{Fig2SM}.
}
\label{f:twolink}
\end{figure}
First, we show that such an object cannot generate locomotion when it is
deformed slowly.
The swimmer is deformed periodically as
\begin{equation} \label{chi1}
\chi_1(t) = A \left( 1 - \frac{1}{2} \sin^2 \frac{\omega t}{2} \right),
\end{equation}
where $A / 2$ is the amplitude and $\omega$ is a frequency of the
oscillation.
Figure~\ref{f:twolink} shows the time evolution of the system for $A = \pi
/ 4$ and $\omega = 0.025 \tau^{-1}$.
The swimmer is initially formed by two elliptic bodies that are joined at
an angle $\pi / 2$, and then it closes and opens periodically, as shown in
Fig.~\ref{f:twolink}(a).
We can see in Fig.~\ref{f:twolink}(b) that the action of the swimmer
generates no net locomotion, whereas the position $X$ of the COM moves
back and forth at a frequency $\omega$, while conserving the total
momentum.

The fact that the swimmer cannot propel itself in Fig.~\ref{f:twolink}
reminds us of Purcell's ``scallop theorem''~\cite{Purcell}, which
states that when the motion of a swimmer has time-reversal
symmetry~\cite{Note1}, there is no net locomotion of the swimmer in a
viscous Stokes fluid.
The periodic motion in Eq.~(\ref{chi1}) has time-reversal symmetry, and
therefore the swimmer cannot swim in a Stokes fluid.
Reference~\citen{Chambrion10} showed that the scallop theorem can be
generalized to a perfect fluid, i.e., a periodic motion with time
reversal symmetry never generates locomotion in a perfect fluid.
When the motion of the swimmer is sufficiently slow, and no waves or
quantized vortices are excited, a superfluid can be regarded as a perfect
fluid.
The generalized scallop theorem is therefore applicable to the situation
shown in Fig.~\ref{f:twolink}.
Thus, self-propulsion in a superfluid is impossible for a two-body swimmer
with sufficiently slow deformation.

\begin{figure}
\begin{center}
\includegraphics[width=8.5cm]{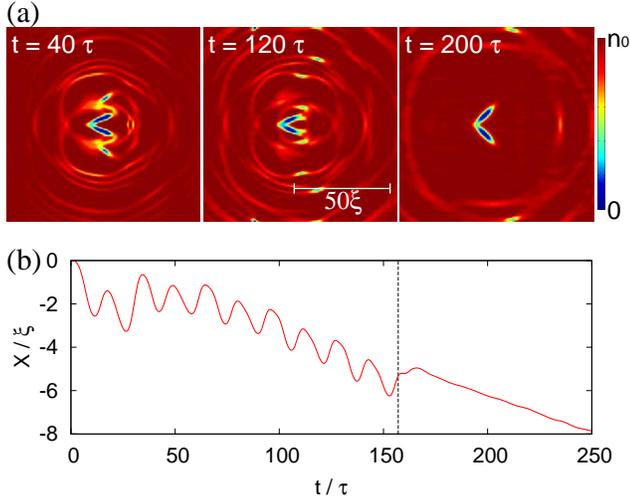}
\end{center}
\caption{
(Color online) Time evolution of (a) the density profile $|\psi|^2$, and
(b) the position $X$ of the COM for $\omega = 0.4 \tau^{-1}$.
The angle $\chi_1(t)$ is driven as in Eq.~(\ref{chi1}) for $\omega t < 20
\pi$, and $\chi_1(t) = 0$ for $\omega t > 20 \pi$.
The dashed line in (b) indicates $\omega t = 20 \pi$.
See the Supplemental Material for a movie of the dynamics shown in
(a)~\cite{Fig3SM}.
}
\label{f:twolink2}
\end{figure}
In the dynamics shown in Fig.~\ref{f:twolink}, the typical velocity of the
elliptic bodies is $\omega d = 0.175 v_s$, and this is much smaller than
the velocity of sound $v_s$; thus, the fluid around the swimmer is barely
disturbed.
When the swimmer is deformed more quickly, on the other hand, waves and
quantized vortices are created, and the condensate can no longer be
regarded as a perfect fluid;
the generalized scallop theorem is no longer applicable.
Figure~\ref{f:twolink2} shows the dynamics of the system for $\omega =
0.4 \tau^{-1}$, which is 16 times larger than the value used for
Fig.~\ref{f:twolink}.
In this case, the shape of the swimmer changes faster than the sound
velocity, and waves and solitonic pulses~\cite{Jones,Berloff} are emitted
from the swimmer, as shown in Fig.~\ref{f:twolink2}(a).
By disturbing the fluid, the swimmer gains net locomotion, as shown in
Fig.~\ref{f:twolink2}(b).
At $\omega t = 20 \pi$ [the vertical dashed line in
Fig.~\ref{f:twolink2}(b)], i.e., after ten strokes, the deformation of the
swimmer is stopped: $\chi_1(t) = 0$ for $\omega t > 20 \pi$.
Even after that, the swimmer continues to travel, which indicates that the
swimmer acquires momentum by emitting waves and solitonic excitations into
the surrounding fluid.

\subsection{Three-body swimmer}
\label{s:3link}

We next study the dynamics of a swimmer consisting of three elliptic
bodies connected by two joints, as illustrated in
Fig.~\ref{f:swimmers}(b).
The Lagrangian in Eq.~(\ref{L}) can be written as
\begin{eqnarray} \label{L3body}
L & = & \frac{3M}{2} (\dot{X}^2 + \dot{Y}^2) + \frac{M d^2}{3} \Bigl[
(\dot{\theta} + \dot{\chi}_1)^2 + (\dot{\theta} + \dot{\chi}_2)^2
\nonumber \\
& & + (\dot{\theta} + \dot{\chi}_1) (\dot{\theta} + \dot{\chi}_2)
\cos(\chi_1 - \chi_2)
+ 3 \dot{\theta}(\dot{\theta} + \dot{\chi}_1) \cos\chi_1
\nonumber \\
& & + 3 \dot{\theta}(\dot{\theta} + \dot{\chi}_2) \cos\chi_2 + 3
\dot{\theta}^2 \Bigr]
\nonumber \\
& & + \frac{I}{2} \left[ \dot{\theta}^2 + (\dot{\theta} + \dot{\chi}_1)^2
+ (\dot{\theta} + \dot{\chi}_2)^2 \right]
\nonumber \\
& & - U_{\rm pot}(X, Y, \theta, \chi_1, \chi_2; \psi),
\end{eqnarray}
which gives the Euler-Lagrange equations,
\begin{subequations}
\begin{eqnarray}
3 M \ddot{X} & = & -\frac{\partial U_{\rm pot}}{\partial X}, \\
3 M \ddot{Y} & = & -\frac{\partial U_{\rm pot}}{\partial Y}, \\
\ddot{\theta} & = & \Biggl\{ -\frac{\partial U_{\rm pot}}{\partial \theta}
-\frac{M d^2}{3} \Bigl[
(\ddot{\chi}_1 + \ddot{\chi}_2) \left(2 + \cos(\chi_1 - \chi_2) \right)
\nonumber \\
& & - (2 \dot{\theta} + \dot{\chi}_1 + \dot{\chi}_2) (\dot{\chi}_1 -
\dot{\chi}_2) \sin(\chi_1 - \chi_2) 
\nonumber \\
& & + 3 \ddot{\chi}_1 \cos\chi_1 + 3 \ddot{\chi}_2 \cos\chi_2 - 3
(2\dot{\theta} + \dot{\chi}_1) \dot{\chi}_1 \sin\chi_1 
\nonumber \\
& & - 3 (2\dot{\theta}
+ \dot{\chi}_2) \dot{\chi}_2 \sin\chi_2 \Bigr]
- I (\ddot{\chi}_1 + \ddot{\chi}_2) \Biggr\}
\nonumber \\
& & / \Biggl\{ \frac{2M d^2}{3} [ 5 + \cos(\chi_1 - \chi_2)
+ 3 \cos\chi_1 
\nonumber \\
& & + 3 \cos\chi_2 ] + 3I \Biggr\}.
\end{eqnarray}
\end{subequations}
There are two controlled parameters ($\chi_1, \chi_2$) that are used to
specify the shape of the swimmer.
When the periodic motion of the swimmer is represented by a closed loop in
$(\chi_1, \chi_2)$ space, the motion is neither time-reversal symmetric
nor reciprocal~\cite{Note1}, and therefore, locomotion is not prohibited
by the generalized scallop theorem.

\begin{figure}
\begin{center}
\includegraphics[width=8.5cm]{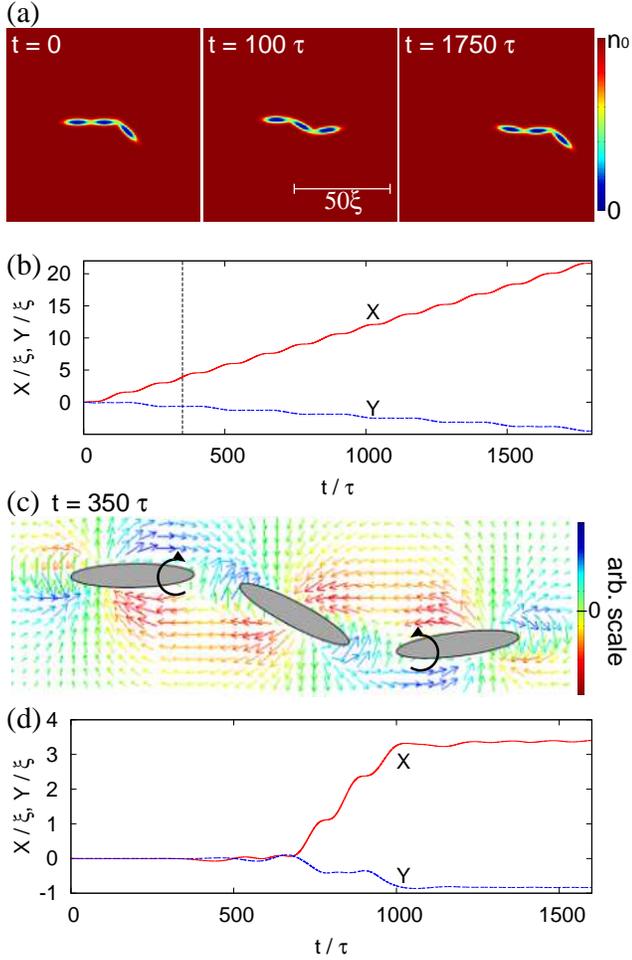}
\end{center}
\caption{
(Color online) Time evolution of (a) the density profile $|\psi|^2$, and
(b) the position of the COM of a three-body swimmer driven as in
Eq.~(\ref{chi12}), with $A = \pi / 4$ and $\gamma = \omega = 0.025
\tau^{-1}$.
See the Supplemental Material for a movie of the dynamics in
(a)~\cite{Fig4SM}.
The vertical line in (b) indicates $t = 350 \tau$.
(c) Flow field $\bm{J}$ at $t = 350 \tau$.
The length of the vector is proportional to $|\bm{J}|$, and the color
represents $J_x$.
The black arrows indicate the directions in which the joints are
rotating.
(d) Time evolution of the position of the COM of the three-body swimmer
driven as in Eq.~(\ref{chi12b}), with $A = \pi / 4$, $\gamma = 2 \times
10^{-5} \tau^{-2}$, $t_0 = 800 \tau$, and $\omega = 0.025 \tau^{-1}$.
}
\label{f:3link}
\end{figure}
Figures~\ref{f:3link}(a) and \ref{f:3link}(b) show the dynamics of the
three-body swimmer for
\begin{subequations} \label{chi12}
\begin{eqnarray}
\chi_1(t) & = & -A \left( 1 - 2 \tanh\gamma t \sin^2 \frac{\omega t}{2}
\right), \\
\chi_2(t) & = & A \tanh\gamma t \sin\omega t,
\end{eqnarray}
\end{subequations}
where $A = \pi/ 4$ and $\gamma = \omega = 0.025 \tau^{-1}$.
The factor $\tanh\gamma t$ is introduced to avoid infinite acceleration
of the body at $t = 0$.
For $t \gg \gamma^{-1}$, Eq.~(\ref{chi12}) reduces to $\chi_1(t) \simeq
-\cos\omega t$ and $\chi_2(t) \simeq \sin\omega t$.
Since the trajectories of $\chi_1(t)$ and $\chi_2(t)$ are circles in
$(\chi_1, \chi_2)$ space, net locomotion is allowed.
We can see from Figs.~\ref{f:3link}(a) and \ref{f:3link}(b) that the
three-body swimmer propels itself without exciting waves or vortices,
in contrast to the two-body swimmer.
The three-body swimmer travels by $\simeq 3 \xi$ for each stroke cycle.

Intuitive understanding of self-propulsion is difficult, because the flow
field
\begin{equation}
\bm{J} = \frac{\hbar}{2mi} \left( \psi^* \nabla \psi - \psi \nabla \psi^*
\right)
\end{equation}
around the three-body swimmer is complicated, as shown in
Fig.~\ref{f:3link}(c).
At $t = 350 \tau$, the position $X$ of the COM of the swimmer is
increasing [vertical line in Fig.~\ref{f:3link}(b)], and hence the
superfluid should have a momentum in the $-x$ direction, due to
conservation of the total momentum.
The red (light gray) region in Fig.~\ref{f:3link}(c) has a large momentum
in the $-x$ direction, and this plays an important role in the locomotion
of the swimmer.

Figure~\ref{f:3link}(d) shows the position of the COM for swimmer's shape
given by
\begin{subequations} \label{chi12b}
\begin{eqnarray}
\chi_1(t) & = & A \left( 1 - 2 e^{-\gamma (t - t_0)^2} \sin^2
\frac{\omega t}{2} \right), \\
\chi_2(t) & = & A e^{-\gamma (t - t_0)^2} \sin\gamma t,
\end{eqnarray}
\end{subequations}
where $A = \pi / 4$, $\gamma = 2 \times 10^{-5} \tau^{-2}$, $t_0 = 800
\tau$, and $\omega = 0.025 \tau^{-1}$.
These functions begin at $\chi_1(0) \simeq \pi / 4$ and $\chi_2(0) \simeq
0$, oscillate while $t / \tau \simeq 800 \pm 300$, and stop at
$\chi_1(\infty) = \pi / 4$ and $\chi_2(\infty) = 0$.
In this case, the swimmer travels only during the deformation, and the
locomotion nearly stops after the shape stops changing.
This behavior is also similar to the self-propulsion of a swimmer in a
viscous Stokes fluid.
The small fluctuations of $X$ and $Y$ for $t / \tau \gtrsim 1200$, shown
in Fig.~\ref{f:3link}(d), are due to a non-adiabatic disturbance of the
condensate.
They vanish when the deformation of the swimmer is sufficiently slow.

\section{Swimming in multicomponent condensates}
\label{s:multi}

We next consider the dynamics of a multicomponent BEC, in which there is a
swimmer who also consists of BECs of different components.
We will consider an immiscible three-component BEC.
Suppose that a swimmer of components 1 and 2 is surrounded by a sea of
component 3.
By controlling the scattering lengths by using the Feshbach resonance, the
size of the droplets of components 1 and 2 can be changed, that is, the
shape of the swimmer can be controlled.

The dynamics of the system can be studied by numerically solving the GP
equation for a three-component BEC
\begin{equation} \label{mGP}
i \hbar \frac{\partial \psi_j}{\partial t} = -\frac{\hbar^2}{2m} \nabla^2
\psi_j + \sum_{k=1}^3 g_{jk} |\psi_k|^2 \psi_j,
\end{equation}
where $j = 1, 2$, and 3.
The interaction coefficients $g_{jk}$ depend on time in such a way that
they always satisfy the immiscible condition: $g_{jk}^2 > g_{jj} g_{kk}$,
where $j \neq k$.
They are changed as
\begin{subequations} \label{g1122}
\begin{eqnarray}
g_{11} & = & \frac{g}{2} \left( 1 + \tanh \gamma t \cos \omega t \right),
\\
g_{22} & = & \frac{g}{2} \left( 1 + \tanh \gamma t \sin \omega t \right),
\end{eqnarray}
\end{subequations}
where $\gamma$ is introduced to avoid an initial disturbance of the system
and $g$ is a parameter that determines the units $\xi$, $v_s$, and
$\tau$.
In the following calculation, we take $\gamma = 0.01 \tau^{-1}$ and
$\omega = 2\pi \times 0.002 \tau^{-1}$.
The other interaction coefficients are fixed as follows: $g_{33} = 1.5 g$
and $g_{12} = g_{23} = g_{13} = g$.
For $t \gg \gamma^{-1}$, the interaction coefficients in Eq.~(\ref{g1122})
become $g_{11} \propto 1 + \cos\omega t$ and $g_{22} \propto 1 +
\sin\omega t$, which do not have time-reversal symmetry.

\begin{figure}
\begin{center}
\includegraphics[width=8.5cm]{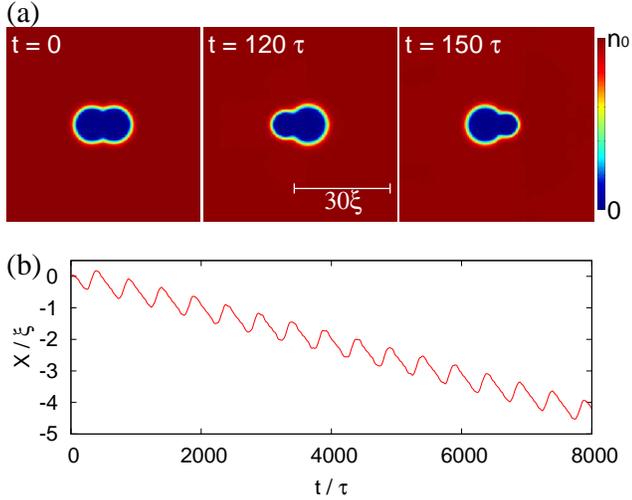}
\end{center}
\caption{
(Color online) Time evolution of (a) the density profile $|\psi_3|^2$ of
component 3, and (b) the position $X$ of the COM of the swimmer.
See the Supplemental Material for a movie of the dynamics shown in
(a)~\cite{Fig5SM}.
}
\label{f:multi}
\end{figure}
The initial state is the ground state with $\int |\psi_1|^2 d\bm{r} = \int
|\psi_2|^2 d\bm{r} = 100 n_0 \xi^2$.
The density $|\psi_3|^2$ of component 3 far from the center is $n_0$, a
constant.
The initial density profile of $|\psi_3|^2$ is shown in the leftmost panel
of Fig.~\ref{f:multi}(a), where the left and right sides of the
low-density region are occupied by components 1 and 2, respectively.
By changing the interaction coefficients as in Eq.~(\ref{g1122}), the
shape of the swimmer deforms, as shown in Fig.~\ref{f:multi}(a).
We find from Fig.~\ref{f:multi}(b) that the swimmer obtains net locomotion
in the $-x$ direction without generating waves or quantized vortices.

The locomotion of the swimmer in Fig.~\ref{f:multi} may be understood
qualitatively as follows.
Suppose that the shapes of components 1 and 2 are approximated by circles
with radii $r_1$ centered at $x_1$ and $r_2$ at $x_2$, respectively,
and that these circles touch each other, i.e.,
\begin{equation} \label{r1r2}
r_1 + r_2 = x_2 - x_1 > 0.
\end{equation}
When component 2 expands or shrinks, component 1 is pushed or pulled as
$\dot x_1 \simeq -\dot r_2$, if the density $n_1$ of component 1 is the
same as the density $n_3$ of the surrounding fluid of component 3.
If $n_1 > n_3$ ($n_1 < n_3$), the displacement of component 1 is
suppressed (enhanced), which may be approximated by $\dot x_1 \simeq -\dot
r_2 + \epsilon (\bar{r} - r_1) \dot r_2$ to the first order of $\bar{r} -
r_1$, where $\bar{r}$ is the radius for which $n_1 = n_3$ and $\epsilon >
0$ is a constant.
The same is also true when component 1 expands or shrinks, and then we
have
\begin{subequations}
\begin{eqnarray}
\dot x_1 & \simeq & -\dot r_2 + \epsilon (\bar{r} - r_1) \dot r_2
- \epsilon (\bar{r} - r_2) \dot r_1, \\
\dot x_2 & \simeq & \dot r_1 - \epsilon (\bar{r} - r_2) \dot r_1
+ \epsilon (\bar{r} - r_1) \dot r_2,
\end{eqnarray}
\end{subequations}
where the third terms on the right-hand side ensure Eq.~(\ref{r1r2}).
The motion of the COM of the swimmer thus becomes
\begin{equation} \label{com}
\dot X = \frac{1}{2}(\dot x_1 + \dot x_2) \simeq
\frac{1}{2} (\dot r_1 - \dot r_2) - \epsilon (\bar{r} - r_2) \dot r_1
+ \epsilon (\bar{r} - r_1) \dot r_2.
\end{equation}
Assuming that $r_1 - \bar{r} \propto \cos\omega t$ and $r_2 - \bar{r}
\propto \sin\omega t$, the time-averaged COM in Eq.~(\ref{com}) becomes
$\bar{\dot{X}} \propto -\epsilon \omega$, which agrees with the fact that
the swimmer travels in the $-x$ direction in Fig.~\ref{f:multi}.
It has been confirmed from numerical simulations of Eq.~(\ref{mGP}) that
the time-averaged velocity $\bar{\dot{X}}$ is proportional to $\omega$, in
agreement with the above result.

The following facts have been confirmed numerically.
In a manner similar to what is shown in Fig.~\ref{f:3link}(c), the
self-propulsion ceases when the swimmer stops deformation.
If the change in the interaction coefficients has time-reversal symmetry,
no net locomotion is obtained, due to the generalized scallop theorem.
For a larger $\omega$, the swimmer emits excitations, in a manner similar
to that shown in Fig.~\ref{f:twolink2}.
However, in this case, components 1 and 2, that is, the body of the
swimmer, are also emitted with waves and solitonic objects of component 3,
and the swimmer gradually shrinks.

\section{Conclusions}
\label{s:conc}

We have investigated the dynamics of deformable objects in superfluids in
order to answer the question: can we swim in a superfluid?
We began by examining the dynamics of articulated bodies, as shown in
Fig.~\ref{f:swimmers}.
For a swimmer that consists of two elliptic bodies, no net locomotion is
achieved when the shape of the swimmer changes sufficiently slowly
(Fig.~\ref{f:twolink}); this is consistent with the generalized scallop
theorem.
When the swimmer is deformed with a velocity comparable to or faster than
the sound velocity, waves and solitonic excitations are emitted into the
superfluid, and the swimmer thus acquires momentum
(Fig.~\ref{f:twolink2}).
A swimmer consisting of three elliptic bodies can swim in a superfluid
without exciting waves (Fig.~\ref{f:3link}), since its motion can break
time-reversal symmetry (non-reciprocal~\cite{Note1}).
We next considered the dynamics of a three-component BEC, in which a
swimmer comprising components 1 and 2 is surrounded by component 3.
The shape of the swimmer can be changed by using the Feshbach resonance to
control the interactions.
We found that it is possible for such a swimmer to create self-propulsion
without exciting waves; this is shown in Fig.~\ref{f:multi}.
Thus, it has been shown that swimming in a superfluid can be done smoothly
or with splashing.

Although, for simplicity, we have considered only uniform 2D systems,
self-propulsion will also be possible in 3D superfluids.
However, spatial inhomogeneity, such as in an atomic gas BEC, may hinder
self-propulsion.
A solution may be possible in a system with ring geometry that has
rotational symmetry instead of translational symmetry;
in such a system, self-propulsion may be verified by a swimmer traveling
around the circumference of the ring.
An interesting area for future research is to consider self-propulsion
methods that are unique to quantum fluids, such as swimming by using
quantized vortices or matter-wave interference.

\begin{acknowledgments}
I wish to thank T. Kiyohara for his participation in an early stage of
this work.
This work was supported by JSPS KAKENHI Grant Number 26400414 and by MEXT
KAKENHI Grant Number 25103007.
\end{acknowledgments}


\begin{thebibliography}{9}

\bibitem{Benjamin}
T. B. Benjamin and A. T. Ellis, Phil. Trans. R. Soc. Lond. A {\bf 260},
221 (1966).

\bibitem{Saffman}
P. G. Saffman, J. Fluid Mech. {\bf 28}, 385 (1967).

\bibitem{Miloh}
T. Miloh and A. Galper, Proc. R. Soc. Lond. A {\bf 442}, 273 (1993).

\bibitem{Kanso}
E. Kanso, J. E. Marsden, C. W. Rowley, and J. B. Melli-Huber, J. Nonlinear
Sci. {\bf 15}, 255 (2005).

\bibitem{Chambrion}
T. Chambrion and A. Munnier, J. Nonlinear Sci. {\bf 21}, 325 (2011).

\bibitem{Donnelly}
R. J. Donnelly, A. N. Karpetis, J. J. Niemela, K. R. Sreenivasan, and
W. F. Vinen, J. Low Temp. Phys. {\bf 126}, 327 (2002).

\bibitem{Zhang}
T. Zhang, D. Celik, and S. W. Van Schiver, J. Low Temp. Phys. {\bf 134},
985 (2004).

\bibitem{Jager}
J. J\"ager, B. Schuderer, and W. Schoepe, Phys. Rev. Lett. {\bf 74},
566 (1995).

\bibitem{Goto}
R. Goto, S. Fujiyama, H. Yano, Y. Nago, H. Hashimoto, K. Obara,
O. Ishikawa, M. Tsubota, and T. Hata, Phys. Rev. Lett. {\bf 100}, 045301
(2008).

\bibitem{Zipkes}
C. Zipkes, S. Palzer, C. Sias, and M. K\"ohl, Nature {\bf 464}, 388
(2010).

\bibitem{Purcell}
E. M. Purcell, Am. J. Phys. {\bf 45}, 3 (1977).

\bibitem{Reeves}
M. T. Reeves, T. P. Billam, B. P. Anderson, and A. S. Bradley,
Phys. Rev. Lett. {\bf 114}, 155302 (2015).

\bibitem{Note1}
More generally, a swimmer cannot obtain locomotion, when its motion is
``reciprocal'', i.e., symmetric with respect to $t \rightarrow -f(t)$ with
$f(t)$ being a monotonically increasing function.
Periodic motion of a swimmer with one degree of freedom, such as in
Fig.~\ref{f:swimmers}(a), is always reciprocal.

\bibitem{Fig2SM}
(Supplemental material) A movie of the dynamics in Fig.~\ref{f:twolink}(a)
is provided online.

\bibitem{Chambrion10}
T. Chambrion and A. Munnier, arXiv:1008.1098.

\bibitem{Fig3SM}
(Supplemental material) A movie of the dynamics in
Fig.~\ref{f:twolink2}(a) is provided online.

\bibitem{Jones}
C. A. Jones and P. H. Roberts, J. Phys. A {\bf 15}, 2599 (1982).

\bibitem{Berloff}
N. G. Berloff, Phys. Rev. B {\bf 65}, 174518 (2002).

\bibitem{Fig4SM}
(Supplemental material) A movie of the dynamics in
Fig.~\ref{f:3link}(a) is provided online.

\bibitem{Fig5SM}
(Supplemental material) A movie of the dynamics in
Fig.~\ref{f:multi}(a) is provided online.

\end{thebibliography}
\end{document}